\documentstyle[11pt,epsfig]{article}

\input epsf

\begin{document}


\title{Buckling Cascade of Thin Plates: Forms, Constraints and Similarity}
\author{B.Roman\footnote{e-mail: roman@lrc.univ-mrs.fr} $\;$ and 
A.Pocheau\footnote{e-mail: pocheau@lrc.univ-mrs.fr} 
\\ IRPHE, UMR 6594 
CNRS, Universit\'es Aix-Marseille I \& II\\
Centre de Saint-J\'er\^ome, S.252, 13397 Marseille Cedex 20, France
} 



\maketitle

\begin{abstract}
We experimentally study compression of thin plates in rectangular boxes 
with variable height.  A cascade of buckling is generated.  It gives 
rise to a self-similar evolution of elastic reaction of plates with 
box height which surprisingly exhibits repetitive vanishing and 
negative stiffness.  These features are understood from properties of 
Euler's equation for elastica.

\end{abstract}


\vskip \baselineskip

 Elastic thin plates submitted to a sufficiently 
large in-plane load are well-known to spontaneously bend due to 
buckling instability \cite{Landau}.  However, this phenomenon, which 
is often quoted as a canonical example of primary bifurcation in 
out-of-equilibrium systems, has been little investigated in the fully 
non-linear regime \cite{Pomeau,PapierFroisse,Boucif,Kramer,Chai}.  
Yet, it provides an appealing opportunity of yielding pattern 
formations within a non-local variational framework - elasticity - 
free of significant noise disturbances.  In addition, it plays an 
essential role in the mechanical resistance of homogenous or layered 
materials and involves important practical implications regarding 
packaging, safety structures or in-load behaviour of plywood or 
film-substrate composites.  Therefore, for both fundamental and 
practical reasons, the non-linear buckling regime of thin plates 
warrants a renewal of interest from physicists.

\begin{figure} [b]
       	\centering
\epsfig {file=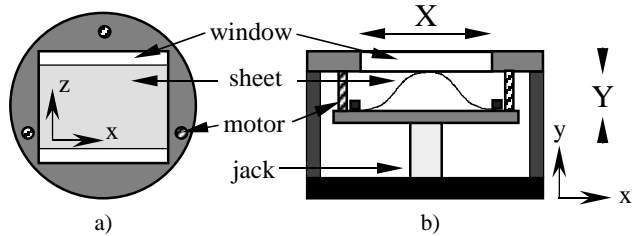,width=8.5cm}
         \caption{ \label{set-up}
        \small {Sketch of the experimental set-up.  a) Top view b) 
       Side view.}}
    \end{figure}

This Letter is devoted to experimentally studying buckling of thin plates from 
the quasi-linear regime to the far non-linear regime.  
In practice, 
plates are confined in a box involving \emph{fixed} horizontal boundaries 
enclosing an area \emph{smaller} than those of plates, and a \emph{variable} 
height $Y$ (Fig.\ref{set-up}).  Plates are thus forced to bend with a 
bending amplitude imposed by the box height.  In 
particular, \emph{iterated} buckling instabilities can then be triggered by 
simply reducing box height.  Doing this, buckled plates display usual 
geometry of out-of-equilibrium patterns: large patches of slightly 
curved folds separated by sharp defects where stretch is mainly localized.

We choose here to put attention on the large patches where bending 
seems dominant.  To this aim, we model them by parallel folds, i.e. 
by \emph{unidirectional} buckling.  This is achieved by taking 
rectangular plates and a rectangular box involving a \emph{constant} 
length $X$, \emph{smaller} than those of plates.  This way, original 
mechanical behaviours have been shown in a framework simple enough to 
be easily handled.  In particular, the occurence and the nature of the 
instability \emph{cascade} have been simply understood by using only 
mechanical invariants and similarity properties.

Reaction force $F$ of buckled plates on the top and bottom boundaries of the 
confining box shows interesting non-linear features: a puzzling 
repetitive \emph{vanishing} of $F$ as box height $Y$ is reduced and, 
in some states, a surprising \emph{negative} stiffness $- dF/ dY <0$.  
Both emphasize the practical differences between the present system 
and common springs.  More generally, the relation $F(Y)$ is shown 
mainly to  exhibit self-similar behaviour.  Self-similarity is stressed 
by mapping most of the cascade of buckling bifurcations onto a 
definite curve with simple rescaling.  These non-linear properties 
are shown to correspond to those of Euler's equation for elastica.

{\it Experimental set-up:} 
 A thin rectangular sheet with dimensions $l \times L$ ($l< L$) and 
 thickness $h$ is clamped on its smallest sides to a bottom plate so as 
 to display an initial bend of height $Y_1$ (Fig.\ref{set-up}).  This is 
 obtained by placing the clamped ends at a distance $X$ shorter than 
 the sheet length $L$ : $X=22cm$, $L=23.3 cm$, $L/X=1.06$, $l=10.1 cm$, 
 $Y_{1}= 35 mm$.  The bottom plate is pushed by a pneumatic jack 
 towards a fixed parallel upper glass plate.  Parallelism between 
 plates during translation is ensured by three guiding axes.  As the 
 distance between confining plates shrinks, the bent sheet comes in 
 contact with the upper plate and starts compressing.  Its compression is 
 stopped by three stepping motors which prevent the bottom plate from 
 moving closer to the upper plate than a controlled distance $Y$.

 At this stage, the buckled plate is thus constrained into a 
 rectangular box still with a lateral extent $X$ but a reduced height $Y\leq 
 Y_{1}$.  Note that the same elastic state could also have been 
 obtained by keeping the height fixed at an initial value $Y$ and 
 reducing the lateral box size to $X$ starting from above 
 \cite{Boucif,Chai,Chateau}.  However, height reduction has been found 
 more convenient here than width reduction to accurately scan a large 
 part of the instability cascade while keeping within the elastic 
 regime.  Moreover, in contrast with the configuration studied in 
 \cite{Boucif}, plates are not held on two opposite sides so as to avoid 
 two-dimensional folds \cite{Pomeau}. On the other hand, the out-plane $Y$-component of the elastical reaction of 
   sheets on the box has been studied instead of the usual in-plane 
   $X$-component \cite{Boucif,Chai,Chateau}.  This is because the former 
   proved to be more sensitive to buckling bifurcation.

 \begin{figure} [t]

\center{
\epsfig{file=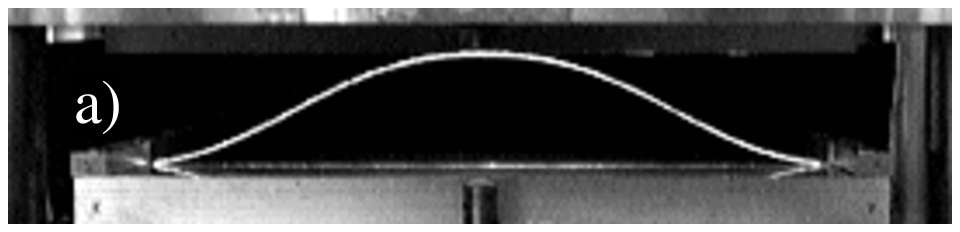,width=4.2cm} 
\epsfig{file=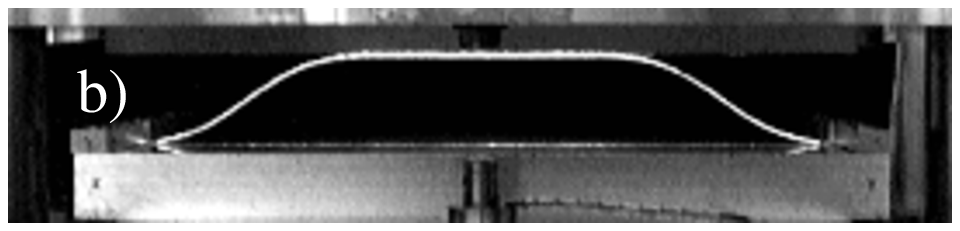,width=4.2cm} }

\center{
\epsfig{file=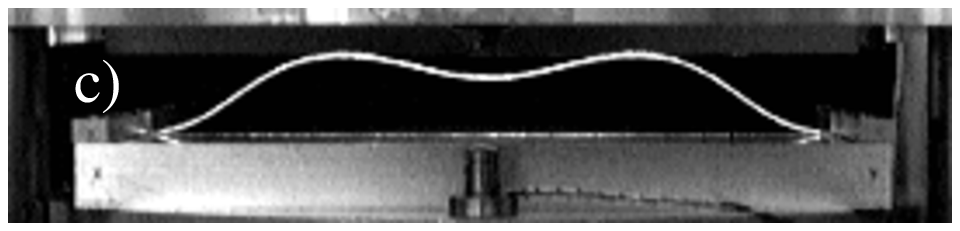,width=4.2cm} 
\epsfig{file=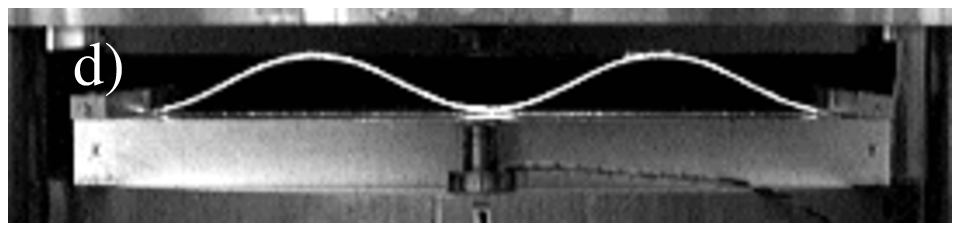,width=4.2cm} }

 \caption
 {\label{ImageSide} \small {Transition from one to two 
 folds: side view of polycarbonate sheet, $h=1mm$, $E\approx 2 GPa$.  
 \emph{Steady} forms a) Contact line: $Y_{1}\!>\!Y\!>\!Y_{1}^{p}$, b) 
 Contact plane: $Y_{1}^{p}>Y>Y_{1}^{b}$, c) Post-buckling with 
 free-standing fold: $Y_{1}^{b}>Y>Y_{2}$, d) Two folds similar to the 
 one of Fig.\ref{ImageSide}a: $Y_{2}>Y>Y_{2}^{p}$.}}

    \end{figure}

\begin{figure} [t]
 
  \center{
\epsfig{file=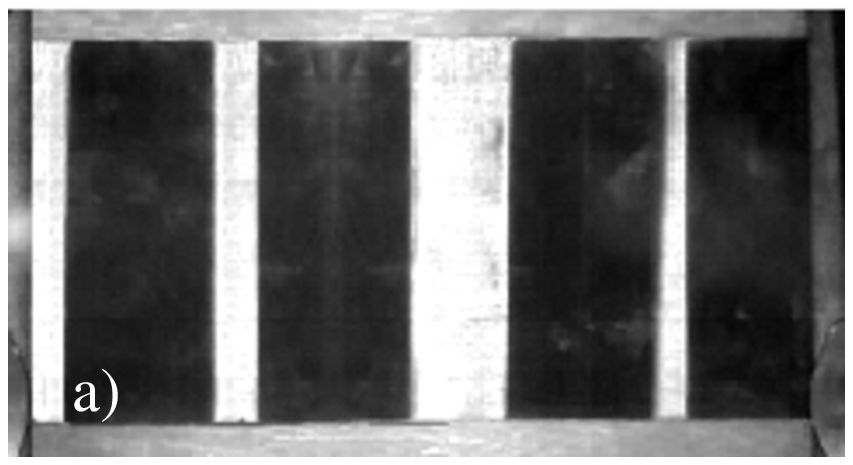,width=4.2cm} 
\epsfig{file=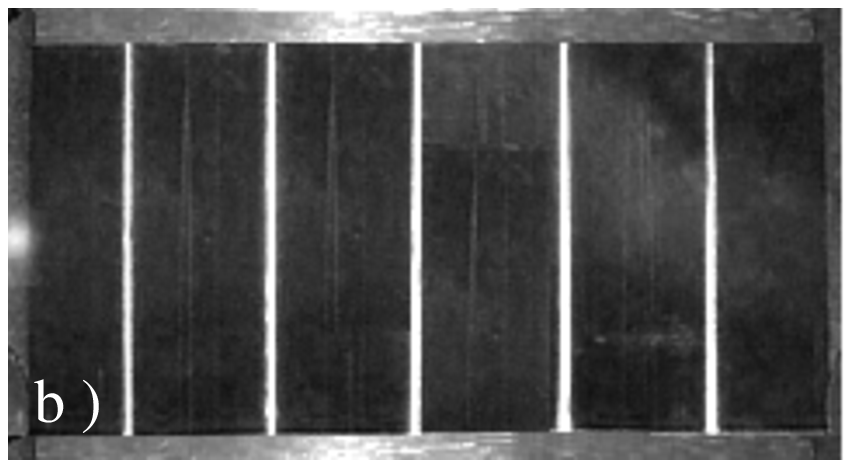,width=4.2cm} }

 \caption 
 {\label{ImageTop} \small {Multi-folds: top view of steel sheet, $E\approx 
 200 GPa$.  a) $h=0.3mm$, contact planes with two folds, b) $h=0.1mm$, 
 contact lines with three folds.}}

\end{figure}

    Height values $Y$ are deduced from stepping motor displacements.
   Forces applied 
   to the jack and to each of the three motors are measured by four load 
   gauges.  The difference $F$ between the former and the latters then 
   gives the normal mechanical response of the sheet.  Sheet forms are 
   observed by light diffusion from the side (Fig.\ref{ImageSide}) or by 
   light reflexion from the top (Fig.\ref{ImageTop}) in which case contacts 
   with either plate appear as bright domains.

{\it Elastic forms:}
All forms described below, either buckled or not, are \emph{steady}. 
Compressing the sheet below the height $Y_{1}$ first gives rise to a 
contact line with the upper plate (Fig.\ref{ImageSide}a).  At a 
critical value $Y_{1}^{p}$, the contact domain becomes a plane through 
a continuous transition.  A measurable part of the sheet is then flat 
and in contact with the upper plate (Fig.\ref{ImageSide}b).  Still 
reducing $Y$, this flat part extends and eventually buckles at another 
threshold $Y_{1}^{b}$.  A new fold is then formed but with a height 
too small to make contact with the other plate (Fig.\ref{ImageSide}c).  
Since no external force is applied at its central point (here, the 
bottom of the fold), we call it a free-standing fold.  As $Y$ is 
further reduced, this fold gets closer to the other plate and 
eventually touches it at a value $Y_{2}$ of $Y$ 
(Fig.\ref{ImageSide}d).  The sheet then exhibits two similar folds in 
contact with both compressing plates.

 When further reducing $Y$, the same evolution resumes: all contact domains 
flatten (Fig.\ref{ImageTop}a) until \emph{one} of them, the 
\emph{largest}, buckles, thereby creating a new fold which eventually 
reaches the other plate at some still lower box height $Y_{3}$ 
(Fig.\ref{ImageTop}b).  This way, an increasing number $n$ of folds 
connecting the upper and bottom plates are created in cascade, at box 
heights $Y_{n}$, all by the same procedure.

\begin{figure}[t]
      	\centering
      	
 \epsfig {file=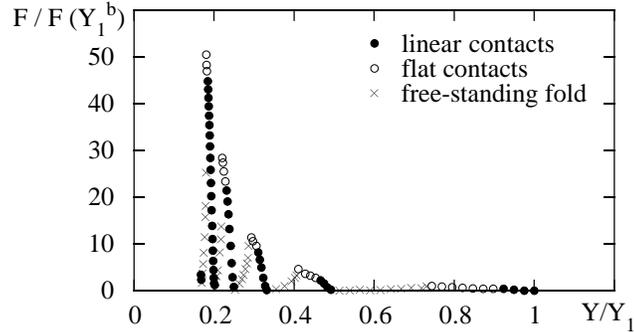,width=8.4cm}
 
  \caption {
  \label{Diagram}
  \small {Reduced reaction force $F/F(Y_{1}^{b})$ on the plates versus 
   reduced fold amplitude $Y/Y_{1}$.  Steel sheets: $h=0.1mm$, $E\approx 200 
  Gpa$, $Y_{1}=35mm$, $F(Y_{1}^{b})=6.06 N$.  Sequence yielding six folds. 
  Linear contacts ($\bullet$), flat contacts ($\circ$), free-standing 
  folds ($\times$).} }
    
    \end{figure}

 {\it Elastic forces:}
 Figure \ref{Diagram} shows, for a typical experiment, a plot of the 
 reduced force $F/F(Y_{1}^{b})$ brought about by a sheet on the set-up 
 plates versus the imposed reduced height $Y/Y_{1}$.  Interestingly, $F$ 
 \emph{vanishes} at several definite box heights.  On the way to such 
 states, the sheet therefore displays elastic forces $F$ 
 \emph{decreasing} with the allowed height $Y$: $dF / dY > 0$.  
 It thus reacts with a larger force when expanded and a smaller force 
 when compressed, thus showing an original elastic behaviour 
 actually opposite to that of springs. In particular, owing to the 
 unusual negative stiffness $- dF / dY < 0$, the sheet should 
 spontaneously collapse under the pressure of the pneumatic jack until 
 a new equilibrium is reached, thereby giving rise to a loss of height 
 control stability.  Such instability of the control parameter $Y$ is 
 fortunately inhibited 
 \noindent here thanks to the reaction of the spacing motors.  A similar 
 decrease of constraint with height has been observed in the somewhat 
 different context of curved strips \cite{Iseki} but only in the weakly 
 non-linear regime.

 The remainder of the letter is devoted to understanding the origin of 
 negative stiffness together with the main properties of the buckling 
 cascade.  This will be done within Euler's equation for thin plates.

{\it Elastica and similarity analysis:}
  Figure \ref{ImageTop} shows a translational invariance of sheet form 
 along the $z$-axis \cite{distribution}.  This property results from the 
 translational invariance of the initial bent state and the homogeneity 
 of box height reduction.  Following it, sheet geometry reduces to that of a 
 curved line: the sheet cut by the (x,y) plane.  We shall parametrize 
 this line by its curvilinear abscissa $s$ and the angle $\theta$ 
 between its tangent and the $x$-axis.  On the other hand, the small 
 thickness of the sheet ($h/L< 1.5 \,10^{-3}$) implies that its flexural 
 rigidity is very weak compared with its extensional rigidity.  
 Sheet is then expected to accomodate height reduction much more by 
 curvature than by extension, except at singularities \cite{PapierFroisse} 
 which will not appear here.
 
 The elasticity problem is thus reduced to an inextensible flexible 
 uniform strip subjected to external loads.  Its equation has been put 
 forward by Euler by expressing mechanical equilibrium of strip 
 elements (\ref{Euler}).  Neglecting gravity effects \cite{Gravity}, 
 force equilibrium shows that the tension force $\textbf{T}$ sustained 
 by strip elements is a constant between contacts with external systems 
 (here top and bottom plates).  Provided that sheet curvature radius 
 $R$ is large with respect to sheet thickness $h$, moments applied to 
 strip elements can be expanded at first order in $1/R$: this 
 corresponds to Hooke's regime.  Then, moment equilibrium yields 
 Euler's equation of elastica \cite{Landau}:
 \begin{equation}
 \label{Euler}
  EI{d^2 \theta \over ds^2} = - p\sin(\theta) + q\cos(\theta)
 \end{equation} 
 Here $E$ is the Young's modulus of the material, $I=lh^3/12$ the 
 inertia moment of a sheet section and $p$ ($q$) the $x$ ($y$)-component
  of tension force $\textbf{T}$.  Boundary conditions are $\theta 
  =0$ at clamped ends.  Global constraints are imposed by the prescribed 
  sheet length $L$ and the box dimensions $(X,Y)$.  The 
  great interest of this equation, as opposed to other elasticity 
  equations, is that it is equally valid for large distortions, provided 
  that sheet curvature radius remains large with respect to sheet 
  thickness.

Our purpose will not be to solve Euler's equation directly but, instead, to 
link its solutions by similarity relations.  On a primary ground, dimensional 
analysis only states that reaction forces on horizontal plates, $F$, and 
lateral sides, $G$, are linked to height $Y$ and sheet length $L$
 by a relationship at fixed 
$X/L$: $FL^{2}/EI=\phi(GL^{2}/EI,Y/L)$.  Exploiting properties of 
elastica, we shall show that, on specific states, $\phi(.,.)$ reduces 
to a monovariate function $\psi(.)$ of variable $Y/L$ exhibiting 
discrete scaling invariance: $\psi(x/n)=n^{3}\psi(x), n\in N$.  This 
will be obtained in two steps: first by building, from a given 
solution in a given box, other solutions in boxes of larger width but 
same height; second, by mapping them to solutions pertaining to a box 
involving the initial width but a smaller height.

Euler's equation (\ref{Euler}) applies to free branches between 
the top and bottom plates.  These branches are connected by 
contacts with plates which will be assumed frictionless here.  Accordingly,  
plates act on the sheet by forces only directed in the $y$-direction 
and with no moment.  In this case, mechanical equilibrium implies that both $p$ 
and $d\theta / ds$ are continuous at the contact points.  Moreover, as 
$p$ is a constant of each free branch, it is therefore 
\emph{invariant} all along the sheet.

These properties enable us to build a solution in larger boxes by 
replication.  Consider a symmetric fold as displayed in the experiment 
(Fig.\ref{set-up}a).  Because of clamping boundary conditions $(\theta=0)$ 
and fold symmetry ($d\theta/ds(s)=d\theta/ds(L-s)$), both $\theta$ 
and $d\theta/ds$ take the \emph{same} values at sheet ends.  
On the other hand, Euler's equation (\ref{Euler}) is \emph{autonomous} 
since all its coefficients are constant on each free branch of the 
sheet.  Thus, pursuing the symmetric fold by itself actually provides 
a relevant solution for elastica.  This means that the curve made by 
two twin symmetric folds indeed corresponds to the form of a 
compressed sheet.  By iterating this procedure, one then gets an 
elastic solution for a sheet of length $nL$ in a box of width $nX$ and 
height $Y$ in term of a chain of $n$ symmetrical folds of length $L$ 
compressed within boxes of width $X$ and height $Y$.

We now use the fact that, as the sheet is thin and its geometry smooth, 
its thickness is not a relevant lengthscale of the elastic problem 
\cite{singular}.  The system thus involves no characteristic scale and 
is therefore scale-invariant.  Accordingly, zooming a shape gives a 
relevant shape for another sheet in another box.  In particular, 
zooming out the previous solution with $n$ folds by a factor $1/n$ 
gives a $n$-folds solution for a sheet of length $L$ compressed in
     \begin{figure}[t]
      \centering
\epsfig {file=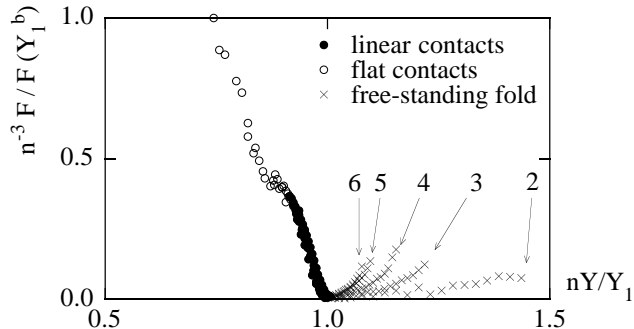,width=8.4cm} \caption {
 \label{ScaledDiagram}
  \small {Rescaling of the bifurcation diagram of Fig.\ref{Diagram} by 
  scaling transformation (\ref{scaling}).  Here, $n$ is the total number 
  of sheet folds, including free-standing folds.  It is indicated for 
  states involving free-standing folds because of lack of dynamic 
  similarity, but would be superfluous for remaining states since 
  scaling (\ref{scaling}) is satisfied.} }
    \end{figure}
 \noindent a box of width $X$ and height $Y/n$.

   Given a symmetric elastic state for an allowed height $Y$, we have 
   thus built a family of elastic states for a series of 
   allowed height $Y/n$.  This result enables us to connect the physical 
   properties pertaining to the stable branches of solutions generated by 
   successive buckling bifurcations \emph{without} solving elastica in detail.  
   In particular, the force $F$ acting on the sheet can now be  easily 
   derived by following its changes during the above transformations.
   
     By definition, 
   fold replication does not change the value of the sheet tension 
   $(p,q)$, since each fold is the same and Euler's equation 
   (\ref{Euler}) is autonomous.  On the other hand, zooming conserves the 
   angle $\theta$ but changes the curvilinear abcissa $s$ into $s/n$.  
   However, according to (\ref{Euler}), this corresponds equivalently to 
   changing $p$ into $n^{2}p$ and $q$ into $n^{2}q$.  This `dynamic 
   similarity' means that the solution of length $L$ in a box of height 
   $Y/n$ and width $X$ is relevant to a tension $(n^{2}p,n^{2}q)$.  
   Finally, as each of the $n$ folds exerts a force $2q$ on the upper 
   plate ($q$ for each contact), one obtains the scaling transformation:
  \begin{eqnarray}
  \label {scaling}
  Y& \rightarrow & n^{-1} Y
   \nonumber \\
 F &\rightarrow& n^3 F
\end{eqnarray}
  
  A direct consequence of this scaling is that force $F$ 
  goes to zero for every $Y_{n}=Y_{1}/n$, since $F$ actually vanishes when 
  the upper plate just touches the sheet ($Y= Y_{1}$, Fig.\ref{set-up}a).  
  Accordingly, for $Y=Y_{n}$, sheet states are the same as if there 
  were {\it no contact} with the upper plate \cite{Plate}.

In the way scaling (\ref{scaling}) has been derived, it applies to a 
family of sheets, each made of a series of identical symmetric folds, 
the folds of different sheets being linked by similarity.  As 
may be directly noticed on 
Figs.\ref{ImageSide}a,\ref{ImageSide}d,\ref{ImageTop}b and 
quantitatively confirmed by rescaling and superposition, these 
conditions are indeed satisfied on states involving 
neither planar contact nor free-standing folds.  Hereafter, they will 
be referred to as `linear contacts'.  To exhibit the expected scaling 
property (\ref{scaling}), we plotted rescaled reduced force 
$n^{-3} F/F(Y_{1}^{b})$ versus rescaled reduced height $n Y/Y_{1}$ on 
Fig.\ref{ScaledDiagram} for the same data as those of 
Fig.\ref{Diagram}.  Excellent collapse is obtained for linear contacts 
($\bullet$), thereby demonstrating the scaling.
 
  Collapse on Fig.\ref{ScaledDiagram} even extends to states involving 
  planar contacts with different lengths ($\circ$) 
  (Fig.\ref{ImageSide}b,\ref{ImageTop}a).  This may be surprising since 
  sheets are then \emph{no longer} made of a series of identical symmetrical 
  folds, as assumed in deriving (\ref{scaling}), so that there is at 
  this stage \emph{no longer} reason for similarity to apply.  However, we show 
  below that, regarding the force $F$, these states may be equally 
  considered to be made of such a series of identical folds.  Accordingly, 
  their reaction force still follow similarity despite their actual 
  form does not.

 When a sheet exhibits a planar contact region, it satisfies $\theta=0$ and 
 $d\theta / ds =0$ at the contact points.  On the other hand, due to 
 integrability of Euler's equation (\ref{Euler}), the quantity 
 $H=EI(d\theta / ds)^{2} +p cos(\theta)+q sin(\theta)$ is conserved 
 along each free branch.  Accordingly, the next contact points found 
 by following the sheet starting from the contact plane also satisfy 
 $d\theta / ds=0$.  By iteration this property extends from fold 
 to fold to \emph{any} contact point of the sheet.  In other words, 
 contacts must be either planar or linear \emph{altogether} 
 \cite{planar}.  On the other hand, planar contacts correspond to 
 `rest' states of the dynamic system (\ref{Euler}).  As equation 
 (\ref{Euler}) is autonomous, their length is thus of no importance for 
 the subsequent motion (i.e.  for the form of the next sheet fold).  
 Accordingly, as long as the total length of planar domains is 
 conserved, one may change the planar domain distribution \emph{without} 
 changing the shape of the free branches and, therefore, the force they 
 exert on plates.  Note also that modification of planar regions do not 
 change the force $F$ by itself since these flat parts correspond to 
 $q=0$ anyway.  Accordingly, for any sheet displaying a flat contact 
 domain, there exists another elastic state made of identical symmetric 
 folds and referring to the same force $F$.  As these states satisfy 
 scaling (\ref{scaling}), this explains that collapse in 
 Fig.\ref{ScaledDiagram} extends to states involving planar contacts.

Whereas the distribution of flat parts has no influence 
on the reaction force $F$, we stress that it parametrizes
buckling's occurence.  This comes from the fact that, as the $x$-component 
$p$ of sheet tension is a constant, the buckling threshold of flat parts 
at given ($X$,$Y$,$L$) solely depends on their size.  Thus, the larger 
a flat part, the larger the height value where buckling occurs.  To 
reduce this dependency on sheet form, only sheets involving a 
\emph{single} flat part have been considered in Figs.\ref{Diagram} and 
\ref{ScaledDiagram}.

In Fig.\ref{ScaledDiagram}, states involving free-standing fold 
($\times$) show no collapse.  
Absence of collapse is due to the fact that a free-standing fold 
cannot be distributed along the sheet, thereby making scaling 
(\ref{scaling}) inapplicable.
On the other hand, branches exhibit a discontinuity with the branch 
relevant to planar contacts.  This simply comes from the step of $n$ at 
each buckling's occurence \cite{n}.

  {\it Conclusion:}
Weakly curved patches of buckled thin plates have been modeled by 
unidirectional buckling for which stretch actually vanishes.  Such 
sheets have been studied in the \emph{far} non-linear domain where 
many buckling bifurcations have already occured.  Then, major elastic 
features such as negative stiffness or vanishing of the normal 
constraint applied to compressing plates can be simply explained from 
similarity considerations and intrinsic properties of Euler's 
equation.  In particular, most of the bifurcation diagram is shown to 
collapse on a curve, thereby stressing the generation of a cascade of 
similar sub-scales as the bifurcation diagram is explored.

On a more general ground, thin plates compressed in rectangular boxes 
 provide an 
elasticity example of cascade generation of self-similar scales in 
out-of-equilibrium systems.  In particular, 
the series of instabilities encountered away from 
equilibrium appear all similar to the primary instability.  
Self-similarity is characterized here by a multiplicative periodicity 
which includes, but goes beyond, log-periodicity.

These results, which are particularly relevant to layered composites, 
show that compressed plates on two perpendicular directions 
alternately turn from a \emph{fragile} to a \emph{robust} state as their 
confinement is increased.  More generally, they show a way of handling 
elasticity without looking directly for elastic solutions.  Finally, 
they provide a useful basis for addressing the more complex case of 
sheets displaying folds in all directions.

  We thank C.Clanet for experimental contribution, J.P. Pahin for 
  preliminary studies, J.Minelli for technical assitance and C.Clanet, 
  M.Abid, T.Frisch for stimulating discussions.


\begin{thebibliography}{99}
\bibitem{distribution} Careful inspection of Fig.\ref{ImageTop}a shows 
a slight width variation of flat parts on the $z$-direction.  This 
follows from the fact that distribution of flat parts is a degree of 
freedom (zero mode) of Euler's equation (\ref{Euler}).
\bibitem{Gravity} Ratio of gravitational ($\rho g h l L Y$) to elastic 
($E I L Y^{2}$) energy is at most $10^{-2}$ here.
\bibitem{singular}Sheet thickness cannot be considered as zero, 
otherwise the factor $EI$ would vanish in (\ref{Euler}).  In this 
sense, it stands as a singular perturbation here.
\bibitem {Plate} Note that this plate is nevertheless necessary to 
sustain folds since, otherwise, all folds would be unstable but the 
one which is clamped at both ends ($Y=Y_{1}$, Fig.\ref{set-up}a).
\bibitem {n} Although discrete change of $n$ is necessary here, 
locating it at the beginning or at the end of buckling is arbitrary: 
it turns out considering a free-standing fold as an actual fold or not.
\end{thebibliography}
\end{document}